\documentclass[twocolumn,pra,showpacs,floatfix,amsmath,amsfonts]{revtex4-1}
\usepackage{multirow}
\usepackage[latin9]{inputenc}
\usepackage{color}
\usepackage{amstext}
\usepackage{graphicx}
\usepackage[table]{xcolor}

\makeatletter
\@ifundefined{textcolor}{}
{%
 \definecolor{BLACK}{gray}{0}
 \definecolor{WHITE}{gray}{1}
 \definecolor{RED}{rgb}{1,0,0}
 \definecolor{GREEN}{rgb}{0,1,0}
 \definecolor{BLUE}{rgb}{0,0,1}
 \definecolor{CYAN}{cmyk}{1,0,0,0}
 \definecolor{MAGENTA}{cmyk}{0,1,0,0}
 \definecolor{YELLOW}{cmyk}{0,0,1,0}
}

\usepackage{color}
\usepackage{graphics}
\usepackage{epsfig}

\newcommand{\R}{\mathbb{R}}

\newcommand{\rs}{\rho_\star}
\newcommand{\ks}{k_\star}
\newcommand{\vphi}{\varphi}
\newcommand{\ve}{{\varepsilon}}

\newcommand{\br}{{\bf r}}

\newcommand{\bB}{{\bf B}}
\newcommand{\bJ}{{\bf J}}
\newcommand{\bE}{{\bf E}}

\newcommand{\p}{{\cal P}}

\newcommand{\cS}{{\cal S}}

\newcommand{\brho}{ \mbox{\boldmath$\rho$\unboldmath}  }

\makeatother

\begin{document}

\title{Linear and nonlinear coherent perfect absorbers on simple layers}

\author{Vladimir V. Konotop$^{1}$ and Dmitry A. Zezyulin$^{2}$}

\affiliation{
 $^{1}$Centro de F\'{i}sica Teórica e Computacional and Departamento de F\'{i}sica, Faculdade de Ciências, Universidade
de Lisboa, Campo Grande, Ed. C8, Lisboa 1749-016, Portugal\\
$^{2}$ITMO University, St.~Petersburg 197101, Russia
}

\date{\today}

\begin{abstract}
 
We consider linear and nonlinear coherent perfect absorbers (CPAs) in multidimensional geometries  and construct explicitly the respective perfectly absorbed solutions. The multidimensional CPAs  have  a structure of the so-called simple layers which represent the generalization of the point $\delta$ function potential to higher dimensions.  The considered examples include broadband CPAs confined to a straight line (in a two-dimensional setting) and to a plane (in the three-dimensional space); CPAs for topological vortices on an absorbing circle; as well as axially-symmetric CPAs on a sphere.  Additionally, it is shown that a paraxial beam propagating along a surface nonlinear CPA embedded in the three-dimensional space can be stable against perturbations. {\color{black} The results are interpreted in applications to optical and acoustic systems.}
 
\end{abstract}

\maketitle

\section{Introduction}
\label{sec:introdcution}

The concept of coherent perfect absorber (CPA) was theoretically introduced in~\cite{StonePRL}, experimentally observed in~\cite{StoneScience}, and received significant attention  over  the last years \cite{review}. A CPA refers to a complex potential which completely absorbs radiation incident on it without any reflection or emission of outgoing waves. It also can be interpreted as a time-reversed laser~\cite{StonePRL}, i.e., as a complex potential which emits radiation in the absence of incident one. 
While the simplest CPAs operate in one-dimensional geometry,  multidimensional CPAs are also known. In the planar 2D geometry, the perfect absorption can be implemented using  composite films of finite width~\cite{film} or lossy ultrathin dielectric layers~\cite{ultra}. Broadband perfect absorption of microwaves by ultrathin films was demonstrated  experimentally~\cite{micro}. Plasmonic perfect absorption of TM-polarized modes by a patterned surface was discussed and implemented experimentally~\cite{patterned}, too. Further results on multi-dimensional CPAs include  the absorption  of electromagnetic waves incident on a spherical barrier~\cite{spherical} and excited inside a 2D cavity with absorbing circular boundary~\cite{cavity}, perfect absorption of TM-polarized modes by a metallings nanocylinders and nanospheres~\cite{cylinder}, exploiting coupling to the surface plasmons. Beyond the optics, cylindrical CPAs for acoustic waves have been reported on in \cite{acoustic}.


Both CPAs and lasers in 1D geometry are characterized by a specific value of the wavelength absorbed and emitted, respectively. Finding such a wavelength requires fine tuning of the parameters. Once the respective wavelength is found and other parameters are fixed, shifting   the wavelength (frequency) from the resonant value destroys the CPA or the laser. In 1D, where the theory can be formulated either in terms of the scattering matrix~\cite{StonePRL} or in terms of the transfer matrix~\cite{Mostafazadeh2009},  the effect of the ``destroying'' a CPA or a laser can be easily understood in terms of the zeros of the respective diagonal elements of the transfer matrix, which are displaced from the real axis under the change of the wavelength or of the system parameters. In 2D and 3D cases, there appears an additional degree of freedom, expressed by the angle of incidence of a beam. Thus even for a monochromatic wave one can pose the question of creating CPAs or lasers operating in a finite-band angular spectrum. Physically, this means that such devices totally absorb or emit radiation at different angles of incidence (emission).

Another aspect of the theory, relevant to the present study, is a possibility of finding exact analytical solutions for  either CPA (or laser) models. When such solutions are available, they allow for quite   complete understanding of the phenomenon. However, even in the 1D setting the problem of spectral singularities (or time-reversal spectral singularities) is not solvable analytically, except for a very few cases, like potentials modeled by the Dirac delta-function~\cite{Mostafazadeh2006}.

In the present paper, we introduce line (in 2D) and surface (in 3D) absorbing or active potentials  which (i) allow for exact analytical solutions,   (ii) some of them operate as   
broadband CPAs (or lasers), (iii) bear Kerr nonlinearity. The corresponding solutions feature totally absorbed incoming radiation without emission (or perfectly emitted radiation without incoming one) in a finite range of the incident angles, although the requirement for the wave to be monochromatic remains.  

Absorption of all incoming radiation 
by linear potentials embedded in nonlinear media  
was reported previously for 
effectively 1D~\cite{Ott} and 2D~\cite{Roman} Bose-Einstein condensates. 
In this work, our interest is in absorbing potentials having either finite or infinite spatial extensions which are either linear or nonlinear and are embedded in a linear medium. In the latter sense, the presented results can be viewed as multidimensional extensions of 1D nonlinear CPAs reported in earlier studies~\cite{nonlinCPA}.  We show that the multidimensional CPA can be made nonlinear, still preserving the above properties of allowing for exact solutions and operating in a finite range of the wavelengths. 

From the mathematical point of view, the potentials considered here are known as {\it simple layers} \cite{Vladimirov} which describe the  Dirac $\delta$ function ``distributed'' on a line or on a surface.  Simple layers  model either absorbing (active) wires incorporated in thin planar waveguides or absorbing (active) surfaces in the three-dimensional space.

The approach employed in this paper is somewhat different from the used in conventional scattering problems, where for given a potential one studies the scattered radiation. We adopt the ``inverse engineering'' approach (similar to one used in~\cite{AKSY} for constructing complex potentials and the respective solutions) where for a given incident wave we construct the distribution of gain-loss landscape, and eventually nonlinear Kerr coefficient of a simple layer ensuring the total absorption or lasing of a given wave.

The rest of the paper is organized as follows. In the next section~\ref{sec:statement} we introduce the main model which is the multidimensional Helmholtz equation with a simple layer potential. In Sec.~\ref{sec:line}   we study the CPAs confined to the lines or curves in a two-dimensional plane. Simple layer CPA for vortices are discussed in Sec.~\ref{sec:vortex}. In Sec.~\ref{sec:parax} we consider the stability of a paraxial beam propagating along the absorbing simple layer. we are interested mainly in optical applications of our results, although in Sec.~\ref{sec:surface} we extend some of the results of the previous sections onto the case of acoustic CPA confined to an absorbing spherical layer. Section~\ref{sec:concl} concludes the paper.

\section{Line and surface potentials}
\label{sec:statement}

We are interested in absorbing or lasing potentials of two geometries: (i) a line (curve) in $\R^2$, 
and (ii) a surface in $\R^3$. Such potentials can be described using  distributions $\delta_S(\br)$, which are referred to as simple layers~\cite{Vladimirov} and represent multidimensional generalizations of the Dirac $\delta$ function. A simple layer $\delta_S(\br)$ is  defined by the following characteristic property:
\begin{eqnarray}
\label{eq:delta}
\int_{\mathbb{R}^D} \delta_S (\br)f(\br)d^D\br=\int_S f(\br)d\cS.
\end{eqnarray} 
Here  $\br\in \R^D$ (in our case $D=2$ or $3$), $f(\br)$ is a  complex-valued test function, and $S$ is the set of   points forming the curve or the surface where the   potential is confined. Respectively, the integral $\int_Sd\cS$ in the r.h.s. of (\ref{eq:delta}) is a curvilinear integral or a surface integral (for the sake of brevity below $S$ is referred to as surface in either case). We will use the property $\delta_S(\br/\ell)=\ell \delta_S(\br)$, which is valid when the dimension of the simple layer is $D-1$. The characteristic scale $\ell>0$ is introduced to make $\delta_S (\br/\ell)$  physically dimensionless.

Let us assume that the surface $S$ is characterized by the permittivity $1+\varepsilon(\br)$ whose variation is complex-valued: $\ve(\br)=\ve_0(\br)+i\gamma(\br)$, where $\br = (x, y, z)$. The parts of the surface with $\gamma(\br)>0$ [$\gamma(\br)<0$] are absorbing [active]. In a general case it will be also  allowed for the layer to bear Kerr nonlinearity described by the real-valued function $\chi(\br)$;  $\chi(\br)>0$ [$\chi(\br)<0$] corresponds to a defocusing [focusing] surface. 

We consider a monochromatic wave, with the frequency $\omega$, incident on the surface   $S$ (from different sides) and  thus governed by the Helmholtz equation:
\begin{equation}
\label{eq:main}
-\nabla^2_\br\psi - k_0^2\delta_S(\br/\ell)[\ve(\br) - \chi(\br)|\psi|^2]\psi =  k_0^2\psi,
\end{equation}
where $\psi$ is one of the components, $E_{x,y,z}$, of the electric field $\bE=(E_x,E_y,E_z)$, $\nabla^2_\br$ is the $3$D  Laplacian, $k_0=\omega/c$, and $c$ is  the speed of light in vacuum. 

  Below, when discussing the electric field, we deal only with the Helmholtz equation (\ref{eq:main}), rather than with the complete system of Maxwell's equations. Thus, {\color{black} strictly speaking,} only the continuity of the tangential components of the electric field will be satisfied, while the field derivatives will have discontinuities at surfaces [these discontinuities are determined by $\ve(\br)$]. Such a wave will have discontinuity of the tangential component of the magnetic field, that is only possible in the presence of the respective surface currents. {\color{black} For a particular example considered in Sec.~\ref{sec:nondif}, this is a current induced in the conducting (absorbing) surface. Respectively, the  obtained surface potential is a CPA for the electromagnetic wave (i.e. for a nondiffractive vectorial field polarized in the plane parallel to the surface).}  

In Sec.~\ref{sec:surface} we will also address Eq.~(\ref{eq:main}) from the point of view of acoustic applications, where $\psi$ will be interpreted as variation of pressure while the simple layer will model an absorbing surface.

\section{Line potentials in a plane}
\label{sec:line}

First we address the situation where a simple layer is $z$-independent, i.e., it can depend only on $x$ and $y$.  
Then, looking for a solution of Eq.~(\ref{eq:main}) in the form of a TE-polarized beam $\bE=(0,0,\psi(x,y))$, we reduce (\ref{eq:main}) to the 2D Helmholtz equation
\begin{equation}
\label{eq:line_main}
-\nabla_{\bot}^2\psi  - k_0^2\delta_S(\brho/\ell)[\varepsilon(\brho) - \chi(\brho)|\psi|^2]\psi = k_0^2\psi
\end{equation}
where $\brho=(x,y)$,  $\nabla_{\bot}=\left(\partial_x,\partial_y\right)$, and the parameter $\ell$ introduced above is chosen as $\ell=1/k_0=\lambda/(2\pi)$, where $\lambda$ is the wavelength of the incident beam.

\subsection{Straight-line CPA}

We start with the simple example when $S$ is a straight line  which coincides with the $x$ axis. Such a potential in (\ref{eq:line_main}) takes the form
\begin{equation}
\label{eq:line}
\delta_S(\brho/\ell)\varepsilon(\brho)= k_0^{-1}\delta(y)\varepsilon(x), 
\end{equation}
where $\delta(y)$ is the conventional Dirac delta function, and $\ve(x)$ describes the variation of the  dielectric permittivity along the line.  Additionally, we assume that the layer is linear, i.e., $\chi(\brho)\equiv 0$.

In order to construct a CPA, i.e., to find distributions of losses (and eventually of gain for construction of a laser) assuring absorption of incident waves for some angles of incidence in the plane $(x,y)$, we look for an incident-beam solution of (\ref{eq:line_main}) in the form
\begin{eqnarray}
\label{ansatz_plane}
\psi_\star(\brho)=\int_{0}^{k_0}\left[\hat{\psi}_+(x,q) +\hat{\psi}_-(x,q)\right]e^{-iq|y|}dq,
\\
\label{ansatz_plane1}
\hat{\psi}_\pm(x,q)=a_\pm(q)e^{\pm i  \sqrt{k_0^2-q^2}x}, 
\end{eqnarray}
where $a_{\pm}(q)$ are, so far,  arbitrary complex  functions describing amplitudes of the angular-spectrum   of the incident beam. 
Obviously, Eq.~(\ref{ansatz_plane}) solves the 2D Helmholtz equation (\ref{eq:line_main})  in the free space, and is continuous at $y=0$, i.e. at  the points where  the line-potential (\ref{eq:line}) is located.  Imposing the  matching condition for the field derivative at $y=0$, 
one can solve the problem as follows.  Given an angular-spectrum distribution of a  beam incident from both sides of the surface, i.e., given  $a_{\pm}(q)$, one can design the dielectric permittivity  landscape $\varepsilon(x)$ which operates as a CPA, i.e.,   absorbs the incident beams entirely. The solution of this problem is given by  
\begin{eqnarray}
\varepsilon(x)=\frac{2i}{k_0}\frac{\int_{0}^{k_0}q\left[\hat{\psi}_+(x,q) +\hat{\psi}_-(x,q)\right]dq}{\int_{0}^{k_0} \left[\hat{\psi}_+(x,q) +\hat{\psi}_-(x,q)\right]dq}
\end{eqnarray}
and is valid for any beam for which the field is different from zero at $y=0$.
 
\begin{figure}
	\includegraphics[width=1.0\columnwidth]{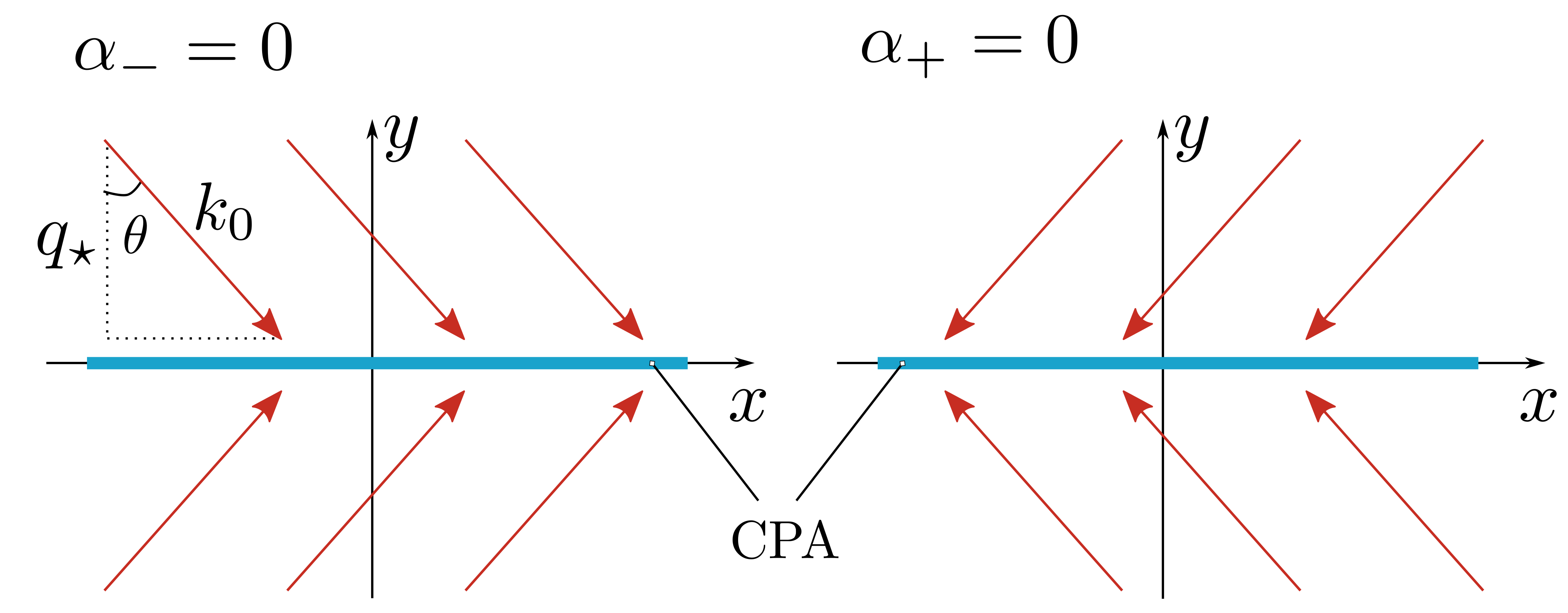}
	\caption{\label{fig:CPAline} 
		Schematic illustration for the  CPA confined in the straight line  and the perfectly absorbed solution (\ref{eq:simple}) for $\alpha_-=0$ (left) and $\alpha_+=0$ (right).}	
\end{figure}

A simple example is given by the CPA for an incident plane wave, as well as for a superposition of plane waves having the same wavelength and equal angles of incidence, $\theta$, the latter defined by $\cos\theta=q/k_0$. Consider $a_\pm(q)=\alpha_\pm\delta(q-k_\star)$, where $k_\star\in(0, k_0)$. Then $\varepsilon(x)\equiv 2ik_\star/k_0$, i.e., $\ve_0 \equiv 0$, and the CPA is characterized by a uniform distribution  of the absorption:  $\gamma(x) \equiv 2 k_\star/k_0$. The corresponding perfectly absorbed solution  reads
\begin{equation}
\label{eq:simple}
\psi_\star(\brho) = e^{-ik_\star|y|}(\alpha_+e^{i\sqrt{k_0^2-k_\star^2}x} +  \alpha_-e^{-i\sqrt{k_0^2-k_\star^2}x}).
\end{equation}
Here $\alpha_\pm$  describe amplitudes of the absorbed waves which propagate in opposite $x$-directions towards the potential as   illustrated in Fig.~\ref{fig:CPAline}.

\subsection{Absorbtion for nondiffracting paraxial beams}
\label{sec:nondif}

The obtained CPA is a straightforward generalization of the 1D result~\cite{Mostafazadeh2006}. It also can be viewed  as {\em broadband} CPA if considered in the context of 3D scattering described by Eq.~(\ref{eq:main}). In this context line (\ref{eq:line}) is nothing but the absorbing plane located at $y=0$. Let us now take into account that $z$-component of the electric field of the form $\psi(x,y)e^{ik_zx}$ {\color{black} formally} solves the Helmholtz equation with $\psi(x,y)$ given by (\ref{ansatz_plane}), (\ref{ansatz_plane1}) with $k_0$ replaced by $\kappa=\sqrt{k_0^2-k_z^2}$.   Indeed, with the constant dielectric permittivity, the potential in model (\ref{eq:main}) becomes separable. The absorption depends only on the $y$--component of the wavevector. Thus any wave with the same $k_\star$ and different $\kappa$, i.e. different $k_z$, will be coherently absorbed [i.e., solution (\ref{eq:simple}) remains valid for such waves], provided $k_\star^2+k_z^2<k_0^2$. 

Thus, a formal solution of (\ref{eq:main}) can be written down as a superposition of four beams ($s_{1}=\pm$ and $s_2=\pm$): 
\begin{equation}
\label{line_general}
\begin{array}{l}
 \psi_{(s_1,s_2)}(\br)=e^{-ik_\star|y|}\times\\[2mm]
 \displaystyle
 \quad\quad\int_{0}^{\sqrt{k_0^2-k_\star^2}}\alpha_{s_1}(k_z)e^{is_1  \sqrt{k_0^2-k_\star^2-k_z^2}x  
  +is_2k_zz}
  dk_z 
 \end{array}
\end{equation} 
where 
$\alpha_{s_1s_2}(k_z)$ describe the angular spectra of the beams. Such   solution will be coherently absorbed by the surface $y=0$ characterized by the absorption 
\begin{eqnarray}
\label{eq:nondiff_surf}
\varepsilon(\br)\equiv 2i\frac{k_\star }{k_0} ={\color{black} i\sigma_\star},
\end{eqnarray}
{\color{black} where $\sigma_\star$ is the surface conductivity.}
Representing $k_z=\sqrt{ k_0^2-k_\star^2}\sin\varphi$ and considering $\varphi$ as an integration variable, we convert (\ref{line_general}) to the form 
\begin{equation}
\label{eq:nondiff1}
\begin{array}{l}
\psi_{(s_1,s_2)}(\br)=e^{-ik_\star|y|}\times\\[2mm]\displaystyle
\quad \quad\quad \int_{0}^{\pi/2}A_{s_1}(\varphi)e^{i\sqrt{k_0^2-k_\star^2}\left(s_1 x\cos\varphi+s_2 z\sin\varphi \right)}
d\varphi,
\end{array}
\end{equation}
where   $A_{s_1}(\varphi) = \sqrt{k_0^2-\ks^2}\cos\vphi\, \alpha_{s_1}(k_z)$. Using a proper  superposition of these four beams,  one can readily construct a general nondiffracting beam \cite{Durnin} 
\begin{equation}
\label{eq:nondiff2}
\psi(\br)=e^{-ik_\star|y|} \int_{0}^{2\pi}A(\varphi)e^{i\sqrt{k_0^2-k_\star^2}\left(x\cos\varphi+z\sin\varphi \right)}
d\varphi.
\end{equation}
In (\ref{eq:nondiff2}), $A(\vphi)$ is a properly defined combinations of amplitudes $A_\pm(\vphi)$.

  Although, expression (\ref{eq:nondiff2}) formally solves  Helmholtz equation (\ref{eq:main}) with the absorbing plane, being interpreted as a component (say, $z$-component) of the electric field, it does not represent the exact solution for the electric field because such field is not a TE wave, and hence the boundary conditions for the field component normal to the surface have to be imposed. Nevertheless, there are at least three cases when formula (\ref{eq:nondiff2}) acquires physical meaning.

First, if $z$-component of the incident field has much larger amplitude than the other two components:  $|E_{x,y}|\ll |E_z|$, then expression (\ref{eq:nondiff2}) becomes the leading order approximation for the field. This, in particular, occurs for the paraxial beams, characterized  by a sufficiently narrow angular spectrum $A(\varphi)$ with the maximum at $\varphi=0,\,\pi$ [notice that the the $\delta$-limit  $A(\varphi)=\delta(\varphi)$ or $A(\varphi)=\delta(\varphi-\pi)$ one recovers the plane wave solutions (\ref{eq:simple})]. In this case, however, the absorbing is not perfect, since it occurs only the $z-$component of the field in the leading approximation.

Second, solution (\ref{eq:nondiff2})  acquires physical meaning for the  pressure field absorbed by a layer (this application is discussed also in Sec.~\ref{sec:surface}). 

Finally, an interesting situation, when the plane is a CPA for nondiffractive beams, occurs when, the beam polarization is in $(x,z)-$plane, i.e. when it is parallel to the absorbing with the absorption defined by (\ref{eq:nondiff_surf}). Nondiffractive solutions in vacuum, can be constructed using the expression of (\ref{eq:nondiff2}) type for $x-$ and $z-$components of the field. Indeed, it is straightforward to verify that the field
\begin{eqnarray}
\label{eq:nondiff}
\bE_\star=e^{-ik_\star|y|} \int_{0}^{2\pi} e^{i\sqrt{k_0^2-k_\star^2}\left(x\cos\varphi+z\sin\varphi \right)}  {\cal A(\varphi)}
\nonumber \\
\times \left[\sin(\varphi)\hat{{\bf i}} -\cos(\varphi)\hat{{\bf k}}\right] 
d\varphi, 
\end{eqnarray} 
where $\cal{A(\varphi)}$ is a scalar function and $\hat{{\bf i}}$ and $\hat{{\bf k}}$ are the unitary vectors along $x$ and $z$ axes, solves Helmholtz equation (\ref{eq:main}) with absorbing plane [and $\chi(\br)\equiv 0$], is polarized in the plane of the absorbing layer, and is divergence free: $\nabla_\br\cdot \bE_\star=0$. 

 The electric field (\ref{eq:nondiff}) satisfies the continuity boundary conditions at $y=0$. However, the respective magnetic field $\bB_\star=1/(ik_0)\nabla\times \bE_\star$ in a general case, has continuous only the normal component, while its  tangential component $\bB_{\star t}$ has discontinuity, which is determined by the discontinuous derivatives of the electric field: 
	\begin{eqnarray}
	\label{eq:nondiff_B}
	\bB_{t\star}= 
	\mbox{sign}(y) \frac{  k_\star}{k_0}  e^{-ik_\star|y|}
	\int_{0}^{2\pi} e^{i\sqrt{k_0^2-k_\star^2}\left(x\cos\varphi+z\sin\varphi \right)}  
	\nonumber \\
	\times  {\cal A(\varphi)} \left[\cos(\varphi)\hat{{\bf i}} +\sin(\varphi)\hat{{\bf k}}\right]	d\varphi. \quad
	\label{eq:B}
 	\end{eqnarray}
\textcolor{black}{Thus in order for the beam (\ref{eq:nondiff}) to be coherently absorbed, there must exist a surface current determined by (\ref{eq:B}). One can ensure that in the case at hand such a  current is induced in the conducting surface by the incident field and is given by the Ohm's law:} 	
\begin{equation}
 \bJ_\star =\hat{\bf j}\times [\bB_{t\star}(y=+0)-\bB_{t\star}(y=-0)]
 =
 {\color{black}
 	\sigma_\star \bE_\star(y=0)},
\end{equation} 
{\color{black} where $\sigma_\star$ is defined by  (\ref{eq:nondiff_surf})}.

\subsection{Line absorber with periodic dielectric permittivity}

A more complex example is the incidence of two plane waves with different $y$--projections of the wavevectors: 
\begin{equation}
a_\pm(q)=\alpha_\pm \delta(q-k_\pm), \quad 0< k_{\pm}<k_0,\quad k_+\neq k_-, 
\end{equation}
on the absorbing potential in a form of a line n 2D. If the real amplitudes of the modes are different, i.e. $\alpha_+\neq\alpha_-$,  one computes the distribution
\begin{eqnarray}
\label{eq:periodic}
\varepsilon (x)=2i\frac{k_+\alpha_+e^{i\sqrt{k_0^2-k_+^2}x} +k_-\alpha_- e^{-i\sqrt{k_0^2-k_-^2}x}}{k_0\alpha_+ e^{i\sqrt{k_0^2-k_+^2}x} +k_0\alpha_-e^{-i\sqrt{k_0^2-k_-^2}x}}
\end{eqnarray}
which is a periodic function of $x$, with the period 
\begin{eqnarray}
L=\frac{2\pi}{ \sqrt{k_0^2-k_+^2}+\sqrt{k_0^2-k_-^2}}. 
\end{eqnarray}
In the limit $|\alpha_+|\to|\alpha_-|$, the distribution $\varepsilon(x)$ becomes singular at the points 
$
x_n=L/2+2\pi n$ (where $n$ is integer).
This is a dimensional effect. It occurs due to the destructive interference of the incident waves propagating in opposite $x-$directions (i.e. parallel the simple layer). Indeed, now the field intensity in the points $(x_n,y)$ becomes zero at nonzero field gradient. Since the discontinuity of the gradient is proportional to the field intensity, it cannot be ``induced'' by any finite potential in the points $x_n$ where the field is zero.

An essential difference from the previous case of the constant dielectric permittivity is that the broken translational invariance along the absorbing line, makes $\varepsilon (x)$ $k_0$-dependent, i.e. its CPA can be realized now only for the given left- and right-incident waves. Now the dielectric permittivity has both real and imaginary parts modulated. An unusual property, is that the described coherent perfect absorption depends on the relation between the complex amplitudes, $\alpha_-/\alpha_+$, rather than on the wavevectors only.

 Potential (\ref{eq:periodic}) has more sophisticate structure, than homogeneously absorbing layers considered in the previous subsections. It requires specific periodic modulation of real and imaginary parts of the dielectric permittivity, with the modulation period larger than the half of the wavelength, $L>\lambda/2$. Since, from the practical point of view, the delta function can be viewed as an idealized limiting model of  a sufficiently narrow dielectric layer, the desired configuration can, in principle, be achieved by doping the dielectric by resonant impurities. Practical implementation of such a simple layer remains a demanding open task.

\subsection{Nonlinear straight-line CPA}

Now we consider the nonlinear problem  (\ref{eq:line}) with a real Kerr nonlinearity,  i.e., we let the CPA-line   bear defocusing nonlinearity with the strength distribution given by the function $\chi_S(\brho)$. 
1D nonlinear CPA were previously studied in~\cite{nonlinCPA}. We again focus on a simple case when the absorber is confined to the straight line $y=0$. Since in the plane ($x,y$) outside the line the field solves the homogeneous Helmholtz equation, we consider incidence of a single plane wave from each side of the absorbing line (rather than beams considered above), i.e., we look for solution of (\ref{eq:line}) in the form  
\begin{eqnarray}
\label{plane}
\psi_\star(\brho)=ae^{is\sqrt{k_0^2-\ks^2}x-i\ks|y|}, \qquad s=\pm 1
\end{eqnarray}
where $a>0$ is the wave amplitude, and the  freedom of choice $s$ reflects the invariance of the model under the inversion of the $x$ axis. Solution (\ref{plane}) is formally tantamount to that obtained for the linear CPA on a line [Eq.~(\ref{eq:simple})] with $\alpha_+=0$ or $\alpha_- = 0$ (Fig.~\ref{fig:CPAline}). However, in the nonlinear case the  matching  condition for the $y-$component of the field gradient  acquires  the form
\begin{eqnarray}
\label{plane_cont_nonlin}
2i\ks  - k_0[\varepsilon(x)+ \chi(x) a^2] =0.
 \end{eqnarray}
 Recalling that the dielectric permittivity of the simple layer is a complex function, and allowing for its real part to vary, $\varepsilon= \varepsilon_0(x)+i\gamma$, we observe that a solution of (\ref{plane_cont_nonlin}) can be found in the case when the  dielectric permittivity and nonlinearity strength are connected as $\ve_0(x) = a^2\chi(x)$. The perfectly absorbed  wave   has the form 
 \begin{eqnarray}
 \label{plane_solut}
 \psi_\star(\brho)= a
 \exp\left( \frac{ is}{2}\sqrt{4- \gamma^2}k_0x  - \frac{i}{2}k_0\gamma|y|\right) 
 \end{eqnarray}
 which is valid for $\varepsilon_0(x)\chi(x)>0$ and sufficiently small losses: $\gamma^2<4$. 

\section{Nonlinear CPA for vortices}
\label{sec:vortex}

All CPAs considered above have infinite extensions. 
Now we consider an example with different properties. We are interested in a spatially limited CPA which has nonzero curvature, and is embedded in the 2D space. More specifically we describe a  CPA in a form of a circle (in 3D this CPA has the cylindrical form) of the radius $\rho_\star>0$ centered at the origin. Concentrating on the simplest case, we assume that the distributions of dielectric permittivity, losses, and nonlinear coefficient are constant on the circle, i.e. $\ve(\br)\equiv\ve$ and $\chi(\br)\equiv\chi$.  In the dimensionless polar coordinates $(\rho,\varphi)$,  where $\rho = \sqrt{x^2 + y^2}$  and $\vphi$ is the polar angle,   Helmholtz equation (\ref{eq:main}), for the electric field in the form $\bE=\psi(\brho)\hat{{\bf k}}$, takes the form
\begin{equation}
-\frac{1}{\rho}\frac{\partial}{\partial \rho}\left(\rho\frac{\partial\psi}{\partial \rho}\right)-\frac{1}{\rho^2}\frac{\partial^2\psi}{\partial\varphi^2}-k_0\delta(\rho-\rho_\star)[\ve - \chi |\psi|^2]\psi=k_0^2\psi.
\end{equation}
We look for solutions carrying integer vorticity (topological charge) $m$: $\psi = e^{im\varphi} R(\rho)$. Then in the free space we have the well known equation 
\begin{equation}
\label{eq:radial}
 R_{\rho\rho} + \rho^{-1}R_\rho + (k_0^2 -   m^2\rho^{-2})R=0,
\end{equation}
 whose general solution is a linear combination of the Bessel functions $J_m(k_0\rho)$ and $Y_m(k_0\rho)$. First, we require the continuity of the field on the circle $S$: $R(\rho_\star-0) = R(\rho_\star+0)$. The second conditions is  the discontinuity of the derivative normal to the simple layer: 
\begin{equation}
\label{eq:disc_R}
  R_\rho(\rho_\star+0) - R_\rho(\rho_\star-0) +k_0 [\ve  - \chi|R(\rho_\star)|^2]R(\rho_\star)=0.
\end{equation}
Note that, generally speaking $R(\rho)$ is a complex valued function, but its amplitude and argument are functions of the radius only. 
  
Using the cross-product  
$$
\p(\rho) = \frac {\pi \rho_\star}{2} [ J_m(k_0\rho_\star)Y_m(k_0\rho) - J_m(k_0\rho)Y_m(k_0\rho_\star)],
$$  the sought solution can be written down as
\begin{equation}
\label{solut_ring}
R(\rho)=C\left\{\begin{array}{ll}
     J_m(k_0\rho)    & \mbox{at $\rho<\rho_\star$}
   \\[2mm]
 \displaystyle{   J_m(k_0\rho) -  \ve k_0 J_m(k_0\rho_\star)\p(\rho)}  &
    \\[2mm]
 \displaystyle{   \quad\quad + |C|^2 \chi k_0 J_m^3(k_0\rho_\star)\p(\rho) } & \mbox{at $\rho>\rho_\star$,} 
\end{array}\right. 
\end{equation}
where  $C$ is  an arbitrary constant.

It is instructive to illustrate the obtained solution using the   quantity $\Pi$ which is proportional to  the radial projection of the  Poynting vector (the proportionality coefficient depends on the physical units used):
\begin{equation}
\label{eq:Poynt}
\Pi(\rho) = i(RR_\rho^* - R^*R_\rho). 
\end{equation}
One immediately observes that $\Pi(\rho)=0$ for $\rho<\rs$, and hence there is no radiation incident over the absorber in the inner domain. For $\rho>\rs$ one computes
\begin{equation}
\Pi(\rho) = - 2|C|^2 \gamma k_0 \frac{\rs}{\rho} J_m^2(k_0 \rho_\star).
\end{equation} 
In the absorbing regime, $\gamma>0$, the radial component of the Poynting vector $\Pi(\rho)$ is negative, which is consistent with the energy transfer from the infinity towards the absorbing surface.  

The found solution however is, in general,  not a CPA solution, since it contains both incident and reflected waves. Indeed, using the well-known asymptotic behavior for Bessel functions:
\begin{equation}
\begin{array}{l}
\displaystyle{J_{m}(k_0\rho ) \to \sqrt{\frac{2}{\pi k_0 \rho}} \cos\left(k_0 \rho -\frac{\pi m}{2} - \frac{\pi}{4}\right),}
\\[2mm]
\displaystyle{Y_{m}(k_0\rho ) \to \sqrt{\frac{2}{\pi k_0 \rho}}  \sin\left(k_0 \rho -\frac{\pi m}{2} - \frac{\pi}{4}\right),}
\end{array}\quad r\to\infty,
\end{equation}
we observe that at large polar radii   solution (\ref{solut_ring}) becomes a linear combination of functions $e^{-ik_0 \rho}/\sqrt{k_0\rho}$ and $e^{ik_0 \rho}/\sqrt{k_0\rho}$ i.e., of incoming and outcoming radiation, respectively.
The CPA regime is characterized by the absence of any reflected radiation. This means that the general solution (\ref{solut_ring}) corresponds to the CPA only if the contribution with $e^{ik_0 \rho}/\sqrt{k_0\rho}$ vanishes. Performing straightforward computation, we obtain that this condition is satisfied if the variation of the  dielectric permittivity is chosen as
 \begin{eqnarray}
 \label{eq:ringCPA}
 \ve_{\star} =\chi |C|^2 J_m^2(k_0 \rs)\qquad\qquad\qquad\qquad\qquad\qquad \nonumber\\ -
 \frac{2 [Y_m(k_0 \rs) -i  J_m(k_0 \rs)]}{\pi k_0 \rs J_m(k_0 \rs)[J_m^2(k_0 \rs)+Y_m^2(k_0 \rs)]}.
 \end{eqnarray}
The imaginary part of the dielectric permittivity is positive reflecting the fact that in the CPA regime the circle is absorbing.  The real part of the dielectric permittivity   can be tuned by the nonlinearity coefficient $\chi$ and by the amplitude of the absorbed solution $C$. On the other hand, relation (\ref{eq:ringCPA})  can be viewed as the definition of the intensity   of the perfectly absorbed nonlinear vortex for a given $\ve_{\star}$, and suitable choice of the nonlinearity $\chi$ ensuring the existence of solution of (\ref{eq:ringCPA}) with respect to  $|C|^2$. Additionally we notice that if $k_0\rs$ is a zero of the Bessel function $J_m(k_0\rs)$, then the expression for $\ve_{\star}$ diverges, which means that the CPA operating at the given wavelengths cannot be implemented using a surface of the corresponding radius.

Substituting (\ref{eq:ringCPA}) to (\ref{solut_ring}), we finally obtain that in the exterior domain of the circle the CPA solution for perfectly absorbed vortices is given by the expression 
\begin{equation}
\label{eq:trueCPA}
R_\star(\rho) = C\, J_m(k_0\rs) H_m^{(2)}(k_0\rho) /  H_m^{(2)}(k_0\rs),
\end{equation}
where $H_m^{(2)} = J_m - iY_m$ is the Hankel function of the second kind.
In the interior domain of the circle, $\rho<\rs$, the solution $R_\star(\rho)$ has the form of the Bessel function  presented in (\ref{solut_ring}). 

In contrast to the CPA solution  on the straight line (\ref{plane_solut}),  the nonlinear CPA solution (\ref{eq:trueCPA}) features the algebraically decaying field amplitude. Inside the absorbing circle, both the  the   incident and reflected radiation  exist, although they are balanced inside the circle (in fact this property is more general and holds for any, either perfectly or not perfectly, absorbed vortex). Outside the circle only the incident radiation exists, which means that the layer operates as a one-sided CPA, in contrast to the nonlinear CPA on the line constructed above in (\ref{plane_solut}), where the radiation is absorbed perfectly from both sides of the line. This property is physically natural, since perfect absorption of the radiation of the cavity inside the circle would require the existence of a source also located inside the cavity. 

\begin{figure}
	\includegraphics[width=1.0\columnwidth]{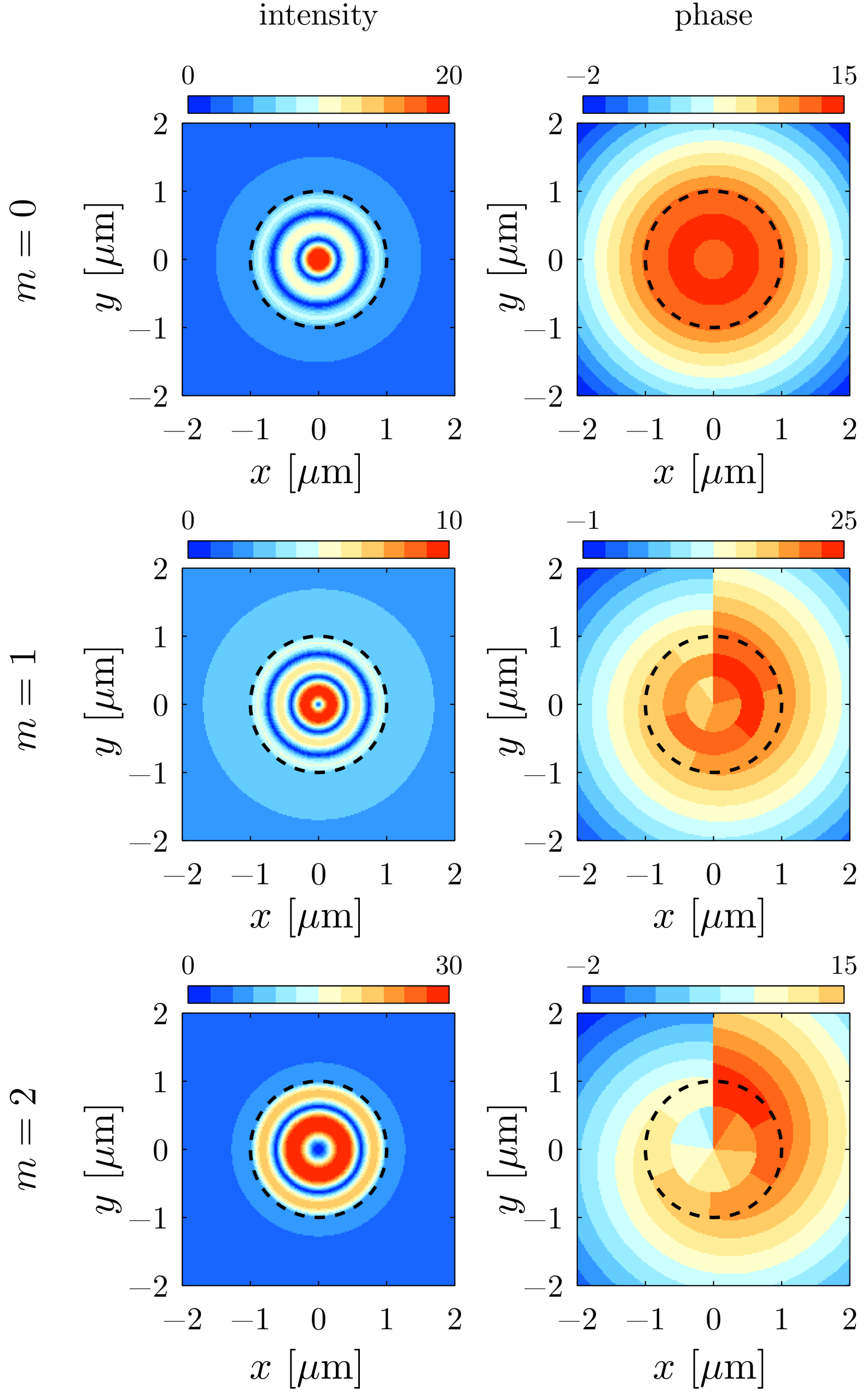}
	\caption{\label{fig:CPAvort} Pseudocolor plots for intensity and phase distributions of  perfectly absorbed nonlinear vortices on an absorbing circular layer for   $\rho_\star=1\,\mu$m  (shown with dashed black line) and the incident wavelength $\lambda=774\,$nm ($m=0,2$) and  $\lambda=667\,$nm ($m=1$). The wavelengths $\lambda$ are chosen to provide the existence condition $Y_m(k_0\rs)/J_m(k_0\rs)>0$. The three rows correspond to (a) $m=0$, $\ve_{\star}\approx 1.00 i$ (b) $m=1$, $\ve_{\star} \approx 1.00i$,  and (c) $m=2$ $\ve_{\star} \approx 0.97i$. In all panels,  $\chi=0.1$.}	
\end{figure}

As a particular example, let us address in more detail the case when the nonlinearity-assisted CPA is implemented using an absorbing layer with the  real part of the  dielectric permittivity fixed to be equal to zero, i.e., $\ve_{\star,0}=0$.  From (\ref{eq:ringCPA}), the corresponding requirement reads
\begin{equation}
\label{eq:sigmaC}
\chi |C|^2 = \frac{2Y_m(k_0 \rs)}{\pi k_0\rs J_m^3(k_0\rs)[J_m^2(k_0\rs) + Y_m^2(k_0\rs)]}.
\end{equation}
If the nonlinearity is defocusing ($\chi >0$) then the nonlinear CPA on such a  layer is possible only if $Y_m(k_0\rs)/J_m(k_0\rs)>0$. 
Examples of 
perfectly absorbed states with topological charges $m=0,1,2$ are presented in Fig.~\ref{fig:CPAvort}.  In this figure, the nonlinearity coefficient is fixed ($\chi=0.1$) and the amplitude of a vortex perfectly absorbed by a purely imaginary layer is computed from (\ref{eq:sigmaC}).

The constructed CPA solution can be easily converted to the solution on a lasing circle. To this end, it is sufficient  to take the complex conjugate of $R_\star(\rho)$ and of $\ve_\star$ (that is to invert sign of imaginary part of the dielectric permittivity replacing the absorbing surface with a lasing one).

\section{Paraxial approximation and stability}
\label{sec:parax}

So far we considered the field distributions in the whole space. Now we turn to the propagation of a paraxial beam along $z$-axis in the half-space $z>0$ in the presence of an absorbing surface defined by a line in the plane ($x,y$) and homogeneous along $z$. Making the standard ansatz $E_z(\br)=\psi(x,y,z)e^{ik_0z}$ with $\psi$ being a slow function of $z$ $|\partial\psi/\partial z|\ll k_0 z$, we obtain the equation of paraxial approximation [recall that $\brho=(x,y)$]: 
\begin{equation}
\label{eq:dyn}
2ik_0\psi_z = - \nabla_{\brho}^2\psi -  k_0^2\delta_S(\brho/\ell)[\varepsilon(\brho) - \chi(\brho)|\psi|^2]\psi.
\end{equation}
On the whole $z$-axis, for the plane wave solution corresponding to $z$-independent amplitude $\psi$ we recover the model (\ref{eq:line_main}).

Now a perfectly absorbed solution, i.e. a beam having no radiation outgoing from the absorbing simple layer, has the form  $\psi(z,\brho) = \psi_\star(\brho) e^{-i\beta z}$, and the question of  its stability (at $z>0$) arises. 
To check the stability of such beam we consider propagation of a perturbed solution in the form $\psi = [\psi_\star + \eta(z,\brho)]e^{-i\beta z}$, where $\xi$ is small perturbation whose behavior  can be described in terms of the linearized version of  equation  (\ref{eq:dyn}):
\begin{equation}
ik_0\eta_\zeta = - \nabla_\rho^2\eta -  k_0^2\delta_S(\brho/\ell)[\varepsilon(\brho)\eta - \chi(\brho)(2|\psi_\star|^2\eta + \psi_\star^2\eta^*)].
\end{equation}
Since the waveguiding medium is linear and the CPA-surface bears defocusing nonlinearity, one can expect that the eventual instability can result only from  spatially localized perturbations. We therefore limit our analysis to perturbations carrying finite power $X(z)=\int_{\mathbb{R}^2}|\eta|^2d\brho$. Direct integration over the whole space $\mathbb{R}^2$ gives 
\begin{equation}
\label{eq:X}
\frac{dX(z)}{dz} = - 2 \int_{S}\gamma(\brho) |\eta|^2d\cS +  2 \textrm{Im\,}\int_S \chi(\brho)\psi_\star^{2}(\eta^{*})^2d\cS.
\end{equation}
If 
\begin{equation}
\label{eq:stab}
\chi(\brho)|\psi(\brho)|^2\leq \gamma(\brho) \quad \mbox{ for all \quad }  \brho\in  S,
\end{equation}
then  the r.h.s. of (\ref{eq:X}) is negative, i.e., the power of perturbations monotonously decays. 

In particular, the stability condition (\ref{eq:stab}) for the CPA solution  on the line \eqref{plane_solut} takes the form $0<\ve(x)\leq \gamma$. For CPA on the nonlinear ring in  Eq.~(\ref{solut_ring}) stability condition  takes the form $\chi |C|^2J_m^2(k_0\rho_\star)\leq \gamma$.

\section{Spherical CPA for acoustic waves}
\label{sec:surface}
{\color{black} Equation (\ref{eq:main}) also describes  interaction of acoustic waves with an absorbing layer, the latter being modeled by the simple layer. Taking into account the  boundary conditions employed in this paper, i.e., the continuity of the field and the discontinuity of the normal derivative, the physical meaning of $\psi(\br)$, in such a statement, is pressure variation field, which is considered dimensionless. Now  $k_0=\omega/c$, where $c$ is the sound velocity in the homogeneous medium, and the sound velocity in the  layer is considered complex and defined by: $c_S^{-2}=c^{-2}[1+\delta_S(\br/\ell)\varepsilon(\br)]$. The  problem for  CPA of acoustic wave can be posed in 3D space for the spherical   simple layer (in 2D setting corresponding to cylindrical absorber  a similar problem was considered in~\cite{acoustic})}. Specifically, we consider an acoustic  CPA on a 
$S$ of a radius $r_\star$ and focus on the simplest  axially symmetric solutions which depend only on the spherical  radius $r =|{\bf r}|$, i.e., on solutions of the form $\psi(\br) = R(r)$. If the sphere is characterized by the uniform distribution of $\varepsilon(\br)=\varepsilon_0+ i\gamma=\,$const,   then the    Helmholtz equation (\ref{eq:main}) reduces to 
\begin{eqnarray}
-\frac{1}{r^2}\frac{d}{dr}\left(r^2\frac{d R}{d r}\right) -  k_0\ve\delta(r-r_\star) R =k_0^2R.
\end{eqnarray}
In the free space the general solution is a linear combination of  functions $\cos(k_0r)/r$ and  $\sin(k_0r)/r$. Imposing the matching conditions at $r=r_\star$, one finds  
\begin{equation}
\label{solut_3Da}
R(r) = \frac{\sin(k_0r)}{k_0r} \qquad\mbox{at $0<r<r_\star$}
\end{equation}
and
\begin{equation}
\label{solut_3Db}
R(r) = \frac{\sin(k_0r)}{k_0r}-\frac{\ve }{k_0r}  {\sin(k_0(r-r_\star))\sin({k_0} r_\star)}
\end{equation}
at $r>r_\star$.
Like in the case of circular absorbing layer, the obtained solution is characterized by the freedom of choice of the coefficients $\varepsilon_0$ and $\gamma$. The radial component of the energy flow can be estimated using the expression $\Pi(r) = i(RR_r^* - R^*R_r)$.  Thus  $\Pi(r)$  vanishes in the cavity inside the  absorbing sphere, while for $r>r_\star$ one computes
\begin{equation}
\Pi(r) = -\frac{2 \gamma \sin^2(k_0 r_\star)}{k_0 r^2},
\end{equation}
which is always negative for $\gamma>0$, thus corresponding to the energy flow from  the infinity towards the absorber.

Since the constructed   solution (\ref{solut_3Db}) is a linear combination of incident ($e^{-ik_0 r}/r$) and reflected ($e^{ik_0 r}/r$) waves, it is, in general, not the perfectly absorbed one. In order to obtain the parameters when the sphere operates as a CPA, we require the outgoing radiation to vanish in the exterior region of the absorber. This requirements leads to the following expression for the variation of the dielectric permittivity:
\begin{equation}
\label{eq:CPA3}
\ve_\star = \cot(k_0 r_\star) + i.
\end{equation}
Substituting (\ref{eq:CPA3}) to the general solution (\ref{solut_3Db}), we obtain that for $r>r_\star$ the perfectly absorbed solution reads
\begin{equation}
R_\star(r) = \frac{\sin(k_0r_\star)}{k_0r}e^{-ik_0(r-r_\star)}.
\end{equation}
Inside the CPA sphere, $r<r_\star$, the solution $R_\star(r)$ is given by expression (\ref{solut_3Da}).

\section{Conclusion}
\label{sec:concl}

In this work, we have constructed explicit solutions for several examples of coherent perfect absorbers (CPAs) confined in straight lines or surfaces embedded in two-dimensional and three-dimensional spaces. From the mathematical point of view, the considered surfaces represent simple layers, which are multidimensional generalizations of the Dirac delta function. We explored different geometries, including line, surface, cylindric and spheric surfaces. Both linear and nonlinear CPAs are presented. Some of the surface CPAs appear to be broadband with respect to the angular spectrum of incidence. The closed lines or surfaces perfectly absorbing vortices do not require incident energy flows from both sides of the surface, and in this sense can be viewed as unidirectional CPAs. Nevertheless, the field is nonzero in the interior domain of the CPA-surface and perfectly absorbed vortices are characterized by the energy flows tangential to the surface either in the interior and exterior domains of the respective CPA. 
Stability of the resulting CPA solutions on nonlinear surfaces has been demonstrated, provided that the intensity  of the solution is below a certain threshold value. 

Finally, the presented coherent perfect absorbers can be straightforwardly transformed in multidimensional simple-layer lasers by changing absorption by gain.

\acknowledgments
 
The research  D.A.Z.  was supported by  Russian Science Foundation (Grant No. 17-11-01004) and by Government of Russian Federation (Grant 08-08).

\end{document}